\documentclass[letterpaper]{article}
\usepackage[preprint]{aaai2027}
\usepackage[hyphens]{url}
\usepackage{graphicx}
\usepackage{natbib}
\usepackage{caption}
\usepackage{booktabs}
\usepackage{array}
\usepackage{tikz}
\usetikzlibrary{arrows.meta,positioning,fit,backgrounds,calc}
\usepackage{amssymb}
\usepackage{docmute}
\renewcommand{\texttt}[1]{\textrm{#1}}
\newcommand{\sgm}{SGM}
\newcommand{\asgm}{ASGM}

\newenvironment{anchoredblock}{%
  \begin{center}\begin{minipage}{\columnwidth}\centering
}{%
  \end{minipage}\end{center}
}

\title{Stigmergic Graph Memory: An Environment-Aware Approach for \\Many-to-Many Multi-Agent Pickup and Delivery}
\author{Aditya Dutta, Joon-Seok Kim$^{*}$}
\affiliations{%
  Emory University\\
  201 Dowman Drive\\
  Atlanta, GA 30322 USA\\
  aditya.dutta@emory.edu, joonseok.kim@emory.edu\\
  $^{*}$Corresponding author
}

\let\arxivbibliography\bibliography
\renewcommand{\bibliography}[1]{}

\begin{document}

\maketitle

\begin{abstract}
Automated fulfillment warehouses must continuously assign and execute pickup-and-delivery work while avoiding congestion. In many-to-many Multi-Agent Pickup and Delivery (MAPD), a request specifies a stock-keeping unit rather than fixed endpoints, requiring the controller to select an agent, source, and destination before path planning. Existing graph-guidance methods primarily influence routing after goals are fixed, leaving endpoint instantiation uninformed by recent traffic. We introduce Stigmergic Graph Memory (SGM), a bounded, decaying memory layer that records recent execution signals on warehouse nodes and directed edges to rank feasible endpoints and route preferences without altering collision constraints or planner validity. Across paired request streams on five layouts, three load levels, and 25 seeds per condition, SGM outperforms two reconstructed many-to-many allocation baselines in all 15 map--load conditions, with paired throughput gains of 20.5--36.7\%. These results show that recent execution memory can improve warehouse throughput by shaping which feasible goals enter the planner, not only how agents travel to already fixed goals.
\end{abstract}

\section{Introduction}
Automated warehouses increasingly rely on mobile-robot fleets for continuous pickup and delivery, making coordination essential for throughput \cite{wurman2008coordinating,boysen2019warehousing}. This problem is commonly formulated as Multi-Agent Pickup and Delivery (MAPD), which combines multi-robot task allocation (MRTA) with collision-free Multi-Agent Path Finding (MAPF) in a shared environment \cite{stern2019mapf,ma2017mapd,xu2022multigoal,korsah2013taxonomy}. The underlying optimization variant of multi-robot path planning is NP-hard \cite{surynek2010optimization,yu2013structure}, making scalable coordination especially important in large, continuously operating warehouses. While most MAPD approaches assume fixed pickup and delivery locations \cite{ma2017mapd,liu2019task}, warehouse requests often identify a stock-keeping unit (SKU) available at multiple sources and deliverable to multiple destinations. Robots must therefore jointly choose an agent, pickup location, and delivery location, yielding a many-to-many MAPD problem.

Recent work on Many-to-Many Multi-Agent Pickup and Delivery (M2M) formulates this setting as a four-dimensional assignment problem over agents, tasks, pickup locations, and delivery locations \cite{schneider2026manytomany}. By exploiting inventory-location flexibility, M2M reports substantial gains over prior approaches, including up to 22,000 additional completed tasks in its eight-hour simulations \cite{schneider2026manytomany}.

However, M2M evaluates assignments before the resulting traffic is realized by the path planner. Its objective relies on estimated travel and inventory-distribution costs and therefore does not directly account for congestion, waiting time, or contention for shared resources. As a result, assignments that appear inexpensive in isolation may repeatedly direct robots toward the same constrained aisles, intersections, or service locations. In dense warehouse environments, these locally attractive decisions can collectively generate traffic patterns that reduce overall system efficiency.

In this paper, we introduce \textbf{Stigmergic Graph Memory (SGM)}, a lightweight memory layer that exposes recent execution patterns to endpoint scoring and routing. Inspired by stigmergic coordination \cite{grass1959stigmergy,theraulaz1999brief}, SGM maintains decaying signals on warehouse nodes and directed edges and maps them to endpoint and route preferences. This lets the controller account for recent traffic when selecting both feasible endpoints and routes.

We compare SGM with reconstructions of M2M and its SKU-aware variant under a common simulator, planner, and paired-request protocol. Across five layouts, three load levels, and 25 seeds per condition, SGM improves throughput in all 15 map--load conditions, with paired gains of 20.5--36.7\%. Routing and matched-cap controls indicate that these gains are not explained by route costs or candidate budgets alone.

We make three contributions. \textbf{(1)} We frame endpoint instantiation in many-to-many MAPD as a congestion-control decision. \textbf{(2)} We introduce SGM, a decaying graph-memory layer that informs endpoint ranking and route preferences. \textbf{(3)} We evaluate SGM through paired benchmarks, routing controls, sensitivity analyses, component ablations, and matched-cap comparisons.

\section{Related Work}

\paragraph{MAPD and many-to-many task allocation.}
Multi-Agent Pickup and Delivery (MAPD) combines multi-robot task allocation (MRTA) with Multi-Agent Path Finding (MAPF) for lifelong task execution. MAPF computes collision-free paths on a graph subject to vertex and edge-conflict constraints \cite{stern2019mapf}, while lifelong MAPF repeatedly assigns new goals and replans over bounded horizons for scalability \cite{li2021lifelong}. MAPD introduces online pickup-and-delivery tasks \cite{ma2017mapd}; related variants include integrated task assignment and path planning \cite{chen2021integrated}, multi-goal MAPD \cite{xu2022multigoal}, and deadline-aware MAPD \cite{wu2021deadlines,makino2024onlineDeadlines}. A broader survey situates MAPD among MAPF, learning-based coordination, and warehouse applications \cite{lau2022mapdreview}.

Classical MAPF methods include reservation-based planning, conflict-based search, safe-interval planning, and scalable priority-based approaches \cite{silver2005cooperative,sharon2015cbs,phillips2011sipp,okumura2022pibt}. Implementation-oriented MAPD variants also account for robot kinematics and continuous execution constraints \cite{honig2016multi,ma2019kinematic}. Integrated MAPD methods further use large-neighborhood search to improve coupled allocation and path-planning decisions \cite{chen2021integrated,ahuja2002vlsn,pisinger2010lns}.

Many-to-many MAPD generalizes fixed pickup-and-delivery tasks by allowing multiple feasible sources and destinations for each SKU request \cite{schneider2026manytomany}. M2M addresses this problem by formulating task allocation as a four-dimensional assignment over agents, tasks, pickup locations, and delivery locations \cite{schneider2026manytomany}, significantly improving throughput over traditional one-to-one MAPD approaches. We therefore use M2M and its SKU-aware variant, M2M-wSKU, as our primary many-to-many MAPD baselines.

\paragraph{Graph guidance and stigmergic coordination.}
Highways, traffic-flow costs, and guidance graphs improve lifelong MAPF by biasing movement toward favorable routes \cite{li2023highway,chen2024trafficflow,zhang2024ggo,zang2025onlineggo,zhang2026mixedggo}. These methods principally affect routing after task goals have been fixed. In contrast, \sgm{} reads \textit{recent execution history} when ranking feasible source--destination (or endpoint) instantiations, which makes our solution unique, as well as when assigning route preferences after endpoints are selected. Table~\ref{tab:related-comparison} summarizes the resulting distinction across four aspects.

\begin{table}[h]
\centering
\caption{Decision interfaces of method families in many-to-many MAPD.}
\label{tab:related-comparison}
\scriptsize
\begin{tabular*}{\columnwidth}{@{\extracolsep{\fill}}lcccc@{}}
\toprule
Family & Source & Dest. & Recent exec. & Routing \\
\midrule
M2M / M2M-wSKU allocation & Yes & Yes & No & No \\
Highways / guidance graphs & No & No & Usually no & Yes \\
Traffic-flow costs & No & No & Sometimes & Yes \\
\sgm{} (ours) & Yes & Yes & Yes & Yes \\
\bottomrule
\end{tabular*}
\end{table}

Stigmergy describes indirect coordination through environmental traces \cite{grass1959stigmergy,theraulaz1999brief}. Ant-colony optimization provides a familiar algorithmic use of decaying environmental traces \cite{dorigo1997ant,bonabeau1999swarm}. \sgm{} adopts stigmergy as a typed, decaying graph-memory abstraction for many-to-many MAPD, rather than as a pheromone-based routing algorithm.

\section{Problem Setting}

We model the warehouse as an undirected graph $G=(V,E)$ with agents $\mathcal{R}=\{a_1,\ldots,a_{\mathcal{I}}\}$. At each discrete timestep, an agent either waits at its current vertex or moves along an adjacent edge. A joint motion plan must satisfy the standard MAPF constraints: no two agents may occupy the same vertex at the same timestep, and no two agents may traverse the same edge in opposite directions during the same timestep.

Tasks arrive online. Each request is $\tau_n=(q_n,\mathcal{S}_n,\mathcal{D}_n,r_n)$, where $q_n$ is a stock-keeping unit (SKU), $\mathcal{S}_n\subseteq V$ is the set of feasible source locations containing $q_n$, $\mathcal{D}_n\subseteq V$ is the set of feasible delivery locations, and $r_n$ is its release time. At time $t$, $\mathcal{T}(t)$ denotes the set of released but uncompleted requests. Instantiating $\tau_n$ selects an agent $a_i$, source $s\in\mathcal{S}_n$, and destination $d\in\mathcal{D}_n$, yielding an assignment $(a_i,\tau_n,s,d)$. The assigned agent must visit $s$ before $d$.

Let $C(H)$ denote the number of tasks completed by the end of planning horizon $H$. The objective is to maximize long-term throughput, $\frac{C(H)}{H}$, subject to task-feasibility and collision-avoidance constraints. The controller must therefore determine task instantiations and collision-free paths over time.

The central challenge is that endpoint instantiation shapes future traffic before path planning begins. A static allocation objective can repeatedly select individually short source--destination pairs that collectively overload the same corridors, storage areas, or service locations. \sgm{} addresses this decision point by using recent execution history to rank feasible endpoint choices before they become planner goals.

\section{Stigmergic Graph Memory}

\begin{figure*}[t]
\centering
\resizebox{0.97\textwidth}{!}{%
\begin{tikzpicture}[
  font=\small,
  >={Stealth[length=2.4mm]},
  node distance=12mm and 13mm,
  box/.style={
    draw, rounded corners=2.5pt, align=center,
    inner sep=5pt, minimum height=13mm, thick
  },
  state/.style={box, fill=black!7, text width=43mm},
  mem/.style={box, fill=orange!12, text width=43mm},
  decision/.style={box, fill=blue!8, text width=37mm},
  planner/.style={box, fill=green!9, text width=40mm},
  exec/.style={box, fill=black!7, text width=40mm},
  memoryflow/.style={->, thick, blue!55!black},
  flow/.style={->, thick, black!72}
]

\node[state] (state) {\textbf{Current Warehouse State}\\[-1pt]
  \footnotesize agent locations, active goals,\\
  \footnotesize released requests};

\node[mem, below=10.4mm of state] (mem) {\textbf{1. Update Execution Memory}\\[2pt]
  \footnotesize \textit{node}: waiting, endpoint pressure, completion\\
  \footnotesize \textit{edge}: traversal, delay, blocking, flow\\[2pt]
  \footnotesize $M_c^{t+1}=\rho_cM_c^t+\Delta_c^t$};

\node[decision, right=13mm of state] (shortlist) {\textbf{2. Shortlist Candidates}\\[2pt]
  \footnotesize feasible agent--source--destination\\
  \footnotesize combinations};

\node[decision, right=11mm of shortlist] (endpoint) {\textbf{3. Endpoint Steering}\\[2pt]
  \footnotesize rank feasible candidates\\
  \footnotesize $C_0+\lambda_pP_{\mathrm{path}}$};

\node[decision, below=13mm of endpoint] (route) {\textbf{4. Route Guidance}\\[2pt]
  \footnotesize derive directed-edge costs\\
  \footnotesize from recent execution memory};

\node[planner, right=16mm of endpoint, yshift=-13mm] (planner) {\textbf{5. RHCR/PBS Planner}\\[2pt]
  \footnotesize selected assignments and route costs\\
  \footnotesize produce joint collision-free plans};

\node[exec, below=12mm of planner] (exec) {\textbf{6. Execute One Timestep}\\[2pt]
  \footnotesize apply planned actions and update\\
  \footnotesize warehouse, pickup, delivery, and completion state};

\draw[flow] (state.east) -- (shortlist.west);
\draw[flow] (shortlist.east) -- (endpoint.west);

\draw[memoryflow] (mem.east) to[out=18,in=-145] (endpoint.south west);
\draw[memoryflow] (mem.east) -- (route.west);

\draw[flow] (endpoint.east) to[out=0,in=150] (planner.west);
\draw[flow] (route.east) to[out=0,in=-150] (planner.west);

\draw[flow] (planner.south) -- (exec.north);

\path (exec.south) ++(0,-7mm) coordinate (feedback-right);
\path (mem.south |- feedback-right) coordinate (feedback-left);

\draw[flow]
  (exec.south) -- (feedback-right) -- (feedback-left) -- (mem.south);

\node[above=1.5mm, font=\footnotesize]
  at ($(feedback-right)!0.5!(feedback-left)$)
  {execution events};

\begin{scope}[on background layer]
  \node[
    draw=blue!45, dashed, rounded corners=3pt,
    fit=(shortlist)(endpoint)(route),
    inner sep=3.8mm
  ] (sgmbox) {};
\end{scope}

\node[
  font=\footnotesize\itshape,
  blue!55!black,
  anchor=south
] at (sgmbox.north) {SGM decision layer};

\end{tikzpicture}%
}
\caption{\sgm{} execution cycle. The current warehouse state provides released requests, agent locations, and active goals. SGM updates typed execution memory, shortlists feasible agent--source--destination candidates, ranks endpoint choices, and derives route costs from recent execution history. RHCR/PBS computes a collision-free joint plan. Executing one timestep updates warehouse state and produces the events used in the next memory update.}\label{fig:arch}
\end{figure*}
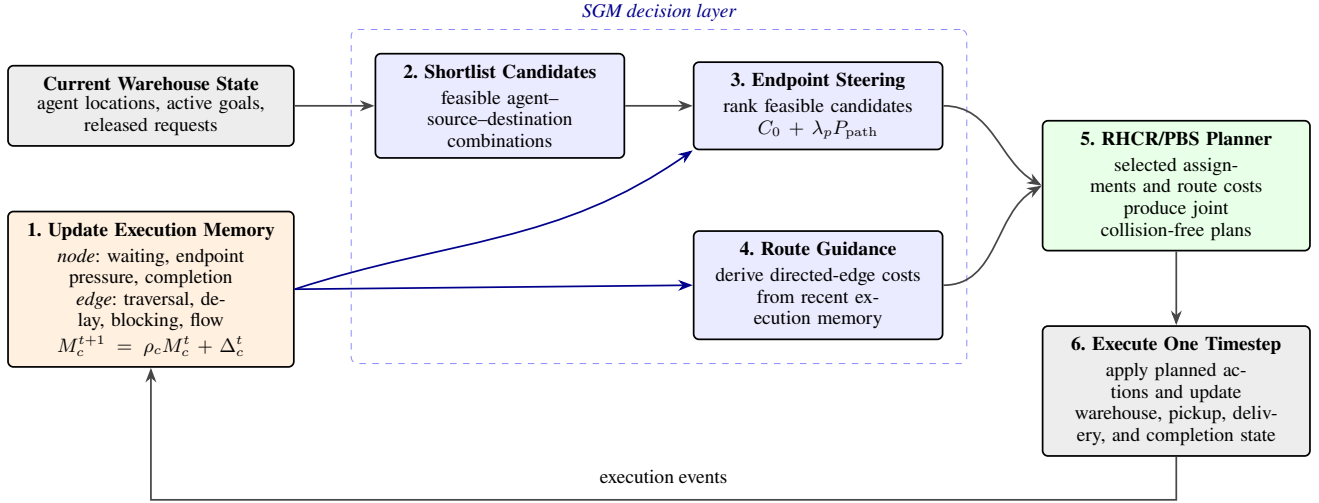

\sgm{} operates within an event-driven rolling-horizon controller based on Rolling-Horizon Collision Resolution (RHCR) \cite{li2021lifelong}, using Priority-Based Search (PBS) as the MAPF backend \cite{ma2019pbs}. Figure~\ref{fig:arch} shows a schematic of \sgm{} architecture. At each control cycle, the system updates memory from recent execution events, instantiates eligible requests, computes collision-free paths for active goals, executes one action from each path, and updates pickup, delivery, and completion states. Replanning occurs when an agent completes a pickup or delivery segment, when a planned path is exhausted, or when execution requires a route repair.

\sgm{} is a shared, decaying memory layer between task instantiation and path planning. It stores separate numerical \textit{channels} for recent node and directed-edge events, which endpoint steering and route guidance read differently.

\subsection{Execution Traces}

\sgm{} does not store a single generic congestion value. Instead, it maintains separate decaying memory channels for different kinds of recent execution information. Node channels track local congestion, endpoint demand, unplanned waiting, pre-execution blocking, and successful task completion. Directed-edge channels track traversal, congestion, delay, pre-execution blocking, and successful directional flow.

Endpoint pressure measures recent demand for a location. When request $\tau_n$ is released, its demand is distributed over feasible endpoints: each source $s\in\mathcal{S}_n$ receives a source-pressure deposit proportional to $1/|\mathcal{S}_n|$, and each destination $d\in\mathcal{D}_n$ receives a destination-pressure deposit proportional to $1/|\mathcal{D}_n|$. Completion memory records successful deliveries associated with an endpoint, while waiting and delay memory record unplanned lack of progress. Planned waits in an otherwise valid route do not create waiting or delay deposits.

Pre-execution blocking pressure is not an executed collision. Before actions are applied, the execution layer checks proposed next moves. If a proposed move would create a vertex conflict or edge swap, the affected agent is held in place and the system records a blocking event at the relevant node and directed edge. The conflicting motion is therefore prevented before execution, so all executed states remain collision-free.

For each channel $c$ and graph element $x$, the corresponding memory value evolves as
\begin{equation}
M_c^{t+1}(x)=\rho_c M_c^t(x)+\Delta_c^t(x),
\label{eq:memory-decay}
\end{equation}
where $\rho_c\in(0,1)$ is the channel-specific retention factor and $\Delta_c^t(x)\geq 0$ is the deposit produced by events at timestep $t$. The empirically derived configuration uses retention values of $0.85$ for waiting and delay, $0.90$ for blocking and endpoint-pressure signals, $0.92$ for congestion, traversal, and flow signals, and $0.95$ for completion signals. An optional corridor adjustment increases retention only up to a cap of $0.99$. Recent events therefore have greater influence than stale events.

Endpoint steering reads endpoint pressure, waiting, and blocking signals, whereas route guidance reads directional traversal, delay, blocking, and flow signals. Keeping these channels separate distinguishes endpoint demand from movement difficulty and directional traffic.

\subsection{Endpoint Steering}

For each feasible many-to-many candidate $(a_i,\tau_n,s,d)$, where $s\in\mathcal{S}_n$ and $d\in\mathcal{D}_n$, \sgm{} starts from the M2M-style baseline allocation cost $C_0(a_i,\tau_n,s,d)$ \cite{schneider2026manytomany}. In the SKU-aware variant, $C_0$ includes the M2M-wSKU inventory-distribution term. \sgm{} then adds a memory-derived path penalty:
\begin{equation}
C(a_i,\tau_n,s,d)
=
C_0(a_i,\tau_n,s,d)
+
\lambda_p P_{\mathrm{path}}(a_i,s,d).
\label{eq:endpoint-score}
\end{equation}

For the reported \sgm{} configuration, the baseline term is $C_0=d(a_i,s)+d(s,d)+C_{\mathrm{SKU}}(\tau_n,s,d)$. The memory term sums current directed-edge penalties along unit-cost shortest-path proxies from the agent to the source and from the source to the destination:
\begin{equation}
P_{\mathrm{path}}(a_i,s,d)
=
\sum_{(u,v)\in \pi(a_i,s)\cup\pi(s,d)}
r_t(u,v),
\label{eq:path-memory}
\end{equation}
where $\pi$ denotes the sampled unit-cost shortest path and $r_t(u,v)=\tilde c_t(u,v)-1$ is the bounded directed-edge penalty defined in Eq.~\ref{eq:movement_cost}. These paths are scoring proxies; RHCR/PBS later computes the final joint plan.

To keep endpoint steering tractable, \sgm{} evaluates SKU-compatible source and destination shortlists capped at 64 locations each, with at most 32 source--destination pairs per request. It can therefore steer among feasible, distance-comparable endpoint choices.

\subsection{Route Guidance}

For routing, \sgm{} converts recent directed-edge memory into bounded positive traversal costs:
\begin{equation}
\tilde c_t(u,v)
=
1+\min\{\kappa,\lambda_r R_t(u,v)\},
\label{eq:movement_cost}
\end{equation}
where $(u,v)$ is a legal directed move, $\lambda_r$ controls the influence of route guidance, and $\kappa$ bounds the maximum increase over the unit free-flow cost. Thus, every legal edge retains positive cost, with $1\leq\tilde c_t(u,v)\leq 1+\kappa$.

The unbounded memory signal is
\[
\begin{array}{@{}r@{}l@{}}
R_t(u,v)=\max\{0,&\alpha_c c_t(v)+\alpha_f f_t(u,v)+\alpha_w w_t(u,v)\\
&+\alpha_b b_t(u,v)-\alpha_s q_t(u,v)+o_t(u,v)\},
\end{array}
\]
where $c_t(v)$ is normalized congestion at the destination node, $f_t(u,v)$ is recent blocking pressure, $w_t(u,v)$ is recent delay, $b_t(u,v)$ measures reverse-direction flow, $q_t(u,v)$ measures successful flow in the forward direction, and $o_t(u,v)$ is the discounted marginal planned-occupancy cost over the guidance horizon. Positive terms discourage recently congested, delayed, blocked, or counter-flow movement, whereas the successful-flow term rewards directions that have recently supported productive traversal.

The reported configuration uses $\lambda_r=0.2$ and $\kappa=0.1$; Supplementary Tables~S3--S4 specify the remaining channel coefficients and controller settings. RHCR \cite{li2021lifelong} uses PBS \cite{ma2019pbs} to compute a collision-free joint plan under these costs.

\subsection{Controller Variants}

A controller determines how tasks are allocated, whether memory is used for endpoint steering and route guidance, when replanning occurs, and how queued work is handled after a route repair. We evaluate two controllers over the same SGM memory representation.

\sgm{} is the primary throughput-oriented controller. At each allocation opportunity, it uses path-memory scoring to rank feasible endpoints and memory-derived route costs when planning active goals. It also preserves queued tasks across route repairs. A \emph{route repair} recomputes paths for active goals after an agent completes a pickup or delivery segment, when an active route is exhausted, or when execution requires replanning. Queue preservation means that unstarted tasks already assigned to an agent remain in that agent's queue rather than returning to the global task pool.

\asgm{} is a guarded, \textbf{A}daptive variant of the same \textbf{SGM} memory layer. Under weak congestion evidence, it uses the M2M-wSKU-style allocation objective and unit-cost routing. As waiting, backlog, blocked movement, or reduced agent availability indicate sustained congestion, it can activate memory-guided endpoint steering and route guidance. Weighted routes are audited against unit-cost alternatives and retained only when they satisfy the guard criteria.

Table~\ref{tab:methods} summarizes the controller policies used in the main benchmark.

\begin{table}[h]
\centering
\caption{Controller policies in the main benchmark.}
\label{tab:methods}
\scriptsize
\begin{tabular*}{\columnwidth}{@{\extracolsep{\fill}}lcccc@{}}
\toprule
Method & SKU & Endpoint memory & Route memory & Queue handling \\
\midrule
M2M & No & No & No & Rebuild \\
M2M-wSKU & Yes & No & No & Rebuild \\
\asgm{} & Yes & Adaptive & Adaptive & Preserve \\
\sgm{} & Yes & Continuous & Continuous & Preserve \\
\bottomrule
\end{tabular*}
\end{table}

\subsection{Validity Boundary}
\label{sec:validity}

\paragraph{Proposition (feasibility preservation).}
For any endpoint assignment selected by \sgm{}, the RHCR/PBS backend receives the original warehouse graph, current agent states, task-feasibility constraints, and collision predicates. Therefore, every joint plan accepted and executed by the backend is collision-free and task-feasible under the original MAPD constraints.

\paragraph{Proof.}
\sgm{} changes only scalar scores used to rank feasible endpoint candidates and bounded positive costs used to rank legal directed-edge traversals. It does not add or remove graph edges, alter agent states, relax endpoint feasibility, or modify vertex- and edge-conflict predicates. RHCR/PBS remains solely responsible for accepting a joint plan. Thus, any executed plan satisfies the original task-feasibility and collision-avoidance constraints.

At endpoint instantiation, memory terms are applied only while ranking a finite set of feasible candidates. At routing, \sgm{} changes only the cost of legal directed moves; for every legal edge $(u,v)$,
\[
1 \leq \tilde c_t(u,v) \leq 1+\kappa.
\]
Because endpoint-memory contributions are clamped and routing costs remain positive and bounded, \sgm{} cannot create infeasible assignments, negative edge costs, zero-cost cycles, or infinite-cost finite paths. It is therefore a bounded preference layer over feasible endpoint and routing decisions.$\square$

\begin{figure*}[t]
\centering
\includegraphics[width=1\textwidth]{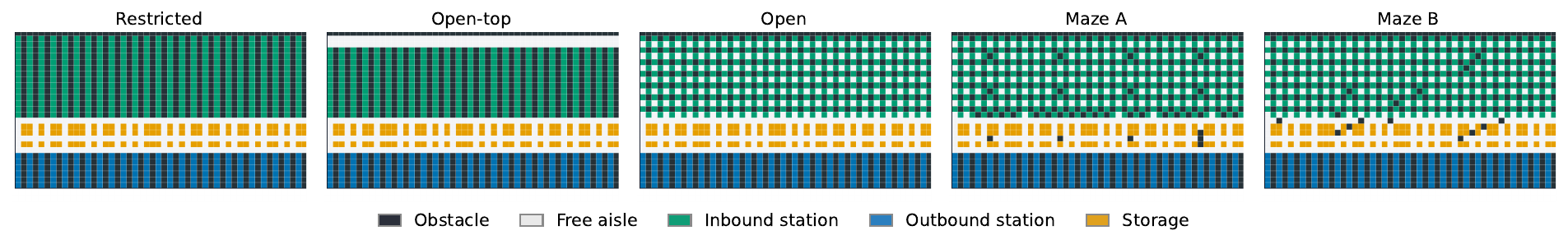}
\caption{The five warehouse layouts, each $50\times27$ cells with 93 designated parking cells. Restricted, open-top, and open are benchmark warehouse layouts from the many-to-many MAPD setting; maze~A and maze~B provide structurally distinct corridor networks at the same scale. Supplementary Table~S5 gives the per-layout cell counts.}
\label{fig:maps-compact}
\end{figure*}

\section{Experimental Protocol}
\label{sec:experimental-protocol}

We evaluate \sgm{} and its adaptive variant, \asgm{}, against M2M and its SKU-aware variant, M2M-wSKU. The main small-map benchmark contains $5\times3\times25\times4=1500$ method--map--fleet-level--seed runs: five warehouse layouts (Figure~\ref{fig:maps-compact}), three parking-normalized fleet levels, and 25 paired seeds.

Each layout contains 93 designated parking cells. We evaluate low, medium, and high fleet levels with 28, 56, and 84 agents, corresponding to 30.1\%, 60.2\%, and 90.3\% of parking capacity, respectively. These fleet sizes do not exceed the number of designated parking endpoints, consistent with the well-formed-infrastructure condition that provides at least one non-task endpoint per agent \cite{ma2017mapd}. Each run spans $H=3600$ timesteps with paired request replay where the step size is one second, so all methods face the same release stream within each map--fleet-level--seed cell.

Let $C(t)$ denote cumulative completed tasks at timestep $t$. We report $C(H)$ as final completed tasks and define horizon throughput as $C(H)/H$. Because every main run uses the same horizon, these measures induce the same method ranking; we use \emph{cumulative completed tasks} for temporal curves and \emph{throughput} only for rates. We also report final-window activity, service time, blocked moves, and allocation cost.

\paragraph{Statistical analysis.}
For each map--fleet-level condition, we compare seed-matched final completed-task counts using two-sided Wilcoxon signed-rank tests \cite{wilcoxon1945individual}. We control the family-wise error rate across the 30 planned comparisons of \sgm{} with M2M and M2M-wSKU using the Holm step-down procedure \cite{holm1979simple}. Paired mean percentage gains and their 95\% confidence intervals are reported as effect sizes; the tests assess whether paired differences are centered at zero. Supplementary Table~S1 gives the complete test record.

\paragraph{Baseline provenance.}
Both baselines are many-to-many allocators reconstructed from the official repository of \citet{schneider2026manytomany}: M2M scores endpoints using travel-distance estimates, and M2M-wSKU adds the inventory-distribution term. We run all methods under the same RHCR/PBS planner, paired request replay, maps, and initial conditions. The baseline reconstructions rebuild queued tasks each timestep, whereas \sgm{} and \asgm{} preserve queued tasks under event-driven replanning. Supplementary Tables~S2 and S4 give the integration and method-specific settings.

We conduct three component ablations over the same $5\times3\times25$ map--fleet-level--seed cells. Endpoint-only retains path-memory endpoint steering but uses unit route costs; routing-only retains memory-weighted routing but removes endpoint-memory scoring; and no-queue-preserve retains both memory interfaces while rebuilding queued tasks each timestep. All ablations use the same paired request streams and RHCR/PBS planning stack as the main methods.

To isolate endpoint steering from routing guidance alone, we evaluate two M2M-wSKU routing controls on the three benchmark warehouse layouts: \emph{M2M-wSKU+recent routing}, which uses SGM recent-traffic edge costs without endpoint memory, and \emph{M2M-wSKU+static highway}, which uses a fixed parity highway. These controls use five paired seeds at the medium and high fleet levels. We also conduct one-factor sensitivity studies, matched-candidate-cap controls, and a five-seed high-fleet transfer study on medium open and open-top layouts. These additional studies support the main 25-seed small-map benchmark rather than replace it.

All main-method comparisons use the same RHCR/PBS planning stack with a planning window of 512, paired request replay, and the shared maps and initial conditions summarized in Table~\ref{tab:methods}. For reproducibility, Supplementary Table~S4 reports method-specific shortlist sizes, endpoint-pair limits, memory weights, guard criteria, and controller settings. Matched-cap controls test whether candidate-search budgets alone explain the observed gains.

\section{Results}

\subsection{SGM improves completed-task throughput}

\begin{table*}[t]
\centering
\caption{Cumulative completed tasks at three checkpoints on the benchmark layouts (mean over 25 paired seeds). \emph{Best M2M} is the stronger of M2M and M2M-wSKU at that checkpoint; its label appears below the value. Parenthetical values for \asgm{} and \sgm{} are percentage differences from that baseline. Bold indicates the highest value among all four methods in a condition--checkpoint cell.}
\label{tab:time-resolved-results}
\scriptsize
\setlength{\tabcolsep}{3.5pt}
\renewcommand{\arraystretch}{0.96}
\newcommand{\rcell}[2]{\shortstack[c]{#1\\[-1pt]{\scriptsize #2}}}
\begin{tabular}{llccccccccc}
\toprule
& &
\multicolumn{3}{c}{Timestep 600} &
\multicolumn{3}{c}{Timestep 1800} &
\multicolumn{3}{c}{Timestep 3600} \\
\cmidrule(lr){3-5}\cmidrule(lr){6-8}\cmidrule(l){9-11}
Layout & Fleet &
Best M2M & \asgm{} & \sgm{} &
Best M2M & \asgm{} & \sgm{} &
Best M2M & \asgm{} & \sgm{} \\
\midrule
Restricted & Low &
\rcell{1140.9}{wSKU} & \rcell{1125.4}{$-1.4\%$} & \rcell{\textbf{1144.6}}{$+0.3\%$} &
\rcell{2736.0}{wSKU} & \rcell{2727.2}{$-0.3\%$} & \rcell{\textbf{3273.6}}{$+19.6\%$} &
\rcell{4610.2}{wSKU} & \rcell{4647.6}{$+0.8\%$} & \rcell{\textbf{6122.7}}{$+32.8\%$} \\

& Medium &
\rcell{1529.0}{wSKU} & \rcell{1566.7}{$+2.5\%$} & \rcell{\textbf{1697.3}}{$+11.0\%$} &
\rcell{3346.6}{wSKU} & \rcell{3412.4}{$+2.0\%$} & \rcell{\textbf{4329.1}}{$+29.4\%$} &
\rcell{5542.2}{wSKU} & \rcell{5628.4}{$+1.6\%$} & \rcell{\textbf{7551.2}}{$+36.2\%$} \\

& High &
\rcell{1808.0}{wSKU} & \rcell{1702.2}{$-5.9\%$} & \rcell{\textbf{1877.8}}{$+3.9\%$} &
\rcell{3796.2}{wSKU} & \rcell{4125.8}{$+8.7\%$} & \rcell{\textbf{4570.5}}{$+20.4\%$} &
\rcell{6280.5}{wSKU} & \rcell{6954.8}{$+10.7\%$} & \rcell{\textbf{7820.6}}{$+24.5\%$} \\
\midrule
Open-top & Low &
\rcell{1091.4}{wSKU} & \rcell{1116.8}{$+2.3\%$} & \rcell{\textbf{1149.2}}{$+5.3\%$} &
\rcell{2639.8}{wSKU} & \rcell{2683.5}{$+1.7\%$} & \rcell{\textbf{3231.2}}{$+22.4\%$} &
\rcell{4313.3}{wSKU} & \rcell{4345.8}{$+0.8\%$} & \rcell{\textbf{5770.4}}{$+33.8\%$} \\

& Medium &
\rcell{1412.6}{wSKU} & \rcell{1502.3}{$+6.4\%$} & \rcell{\textbf{1655.2}}{$+17.2\%$} &
\rcell{3199.8}{wSKU} & \rcell{3323.5}{$+3.9\%$} & \rcell{\textbf{4152.4}}{$+29.8\%$} &
\rcell{5061.9}{M2M} & \rcell{5255.4}{$+3.8\%$} & \rcell{\textbf{6877.9}}{$+35.9\%$} \\

& High &
\rcell{1632.6}{wSKU} & \rcell{1569.1}{$-3.9\%$} & \rcell{\textbf{1809.4}}{$+10.8\%$} &
\rcell{3445.5}{wSKU} & \rcell{3875.5}{$+12.5\%$} & \rcell{\textbf{4393.2}}{$+27.5\%$} &
\rcell{5573.0}{M2M} & \rcell{6156.4}{$+10.5\%$} & \rcell{\textbf{7243.4}}{$+30.0\%$} \\
\midrule
Open & Low &
\rcell{1106.4}{wSKU} & \rcell{1061.9}{$-4.0\%$} & \rcell{\textbf{1111.0}}{$+0.4\%$} &
\rcell{2547.9}{wSKU} & \rcell{2494.8}{$-2.1\%$} & \rcell{\textbf{3076.1}}{$+20.7\%$} &
\rcell{4386.1}{wSKU} & \rcell{4423.4}{$+0.9\%$} & \rcell{\textbf{5820.2}}{$+32.7\%$} \\

& Medium &
\rcell{1468.0}{wSKU} & \rcell{1428.0}{$-2.7\%$} & \rcell{\textbf{1527.1}}{$+4.0\%$} &
\rcell{3156.8}{wSKU} & \rcell{3178.0}{$+0.7\%$} & \rcell{\textbf{3919.6}}{$+24.2\%$} &
\rcell{5269.4}{wSKU} & \rcell{5346.8}{$+1.5\%$} & \rcell{\textbf{6949.2}}{$+31.9\%$} \\

& High &
\rcell{\textbf{1805.2}}{wSKU} & \rcell{1524.1}{$-15.6\%$} & \rcell{1706.9}{$-5.4\%$} &
\rcell{3734.0}{wSKU} & \rcell{3672.0}{$-1.7\%$} & \rcell{\textbf{4165.3}}{$+11.5\%$} &
\rcell{6286.8}{wSKU} & \rcell{6512.2}{$+3.6\%$} & \rcell{\textbf{7516.7}}{$+19.6\%$} \\
\bottomrule
\end{tabular}
\end{table*}

Table~\ref{tab:time-resolved-results} reports cumulative completed tasks at three checkpoints on the restricted, open-top, and open benchmark layouts. \sgm{} is the best method in 26 of the 27 displayed map--fleet--checkpoint cells. The only exception is the earliest checkpoint on the high-fleet open layout, where M2M-wSKU is ahead at timestep 600; \sgm{} overtakes it by timestep 1800 and remains ahead at the end of the horizon. This isolated early deficit reflects \sgm's deliberate warm-up behavior: before sufficient execution events accumulate, its decaying memory exerts limited influence, whereas M2M-wSKU can immediately applies its static SKU-aware allocation objective.

The temporal pattern is consistent across layouts. At timestep 600, \sgm{} is already competitive with, and often exceeds, the stronger M2M baseline. Its larger advantage emerges as traffic accumulates: by timestep 1800, \sgm{} improves on the stronger baseline by roughly 12--29\% in the displayed conditions, and its final gains range from 19.6--36.2\%. Thus, the benefit is not a transient start-up effect. Memory-guided endpoint selection progressively prevents the repeated concentration of work in the same constrained regions.

The full five-layout benchmark gives the same conclusion. Across all 15 map--fleet conditions, \sgm{} exceeds both reconstructed many-to-many baselines under paired request replay, with paired final-throughput gains of 20.5--36.7\%. Supplementary Table~S6 gives the complete five-layout terminal outcomes. The planned paired tests in Section~\ref{sec:experimental-protocol} evaluate these differences under the shared simulator and request streams; Table~\ref{tab:time-resolved-results} shows how the final advantage develops over time rather than only reporting the terminal value.

\subsection{The effect persists across distinct warehouse structures}

\begin{figure}[h]
\centering
\includegraphics[width=1\linewidth]{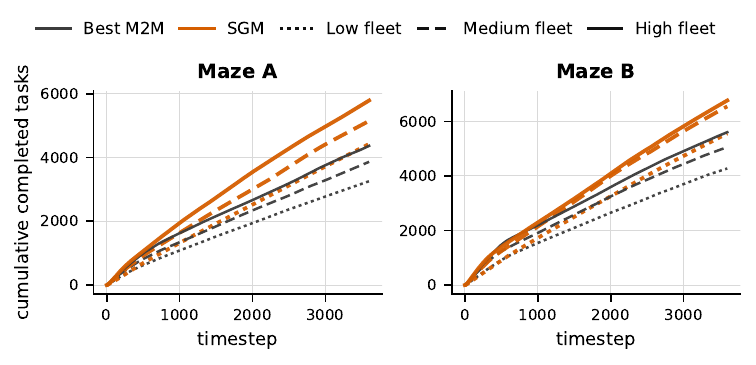}
\caption{Cumulative completed tasks on maze~A and maze~B. Curves average 25 paired seeds. Charcoal denotes the stronger M2M baseline in each map--fleet cell, vermilion denotes \sgm{}, and line style denotes fleet level.}
\label{fig:maze-throughput}
\end{figure}

Figure~\ref{fig:maze-throughput} extends the time-resolved view to the two maze layouts. These maps differ from the benchmark warehouses in their corridor topology and available detours, yet \sgm{} remains ahead of the stronger M2M baseline at medium and high fleet levels.

The routing controls isolate the distinction between learned execution memory and static routing biases. On the three benchmark warehouse layouts at medium and high fleet levels, adding recent-traffic routing alone to M2M-wSKU changes throughput by $-0.1 \pm 2.1\%$, while a fixed parity highway changes it by $-1.4 \pm 2.5\%$. In contrast, \sgm{} improves throughput by $29.2 \pm 5.7\%$ in the same comparison. Routing guidance alone is therefore insufficient: the dominant improvement comes from using recent execution to choose which feasible source--destination pair becomes the task goal.

Additional controls support this interpretation. One-factor sensitivity sweeps retain substantial gains across the examined memory settings, and matched-candidate-cap experiments show that the advantage is not explained by evaluating more endpoint pairs (Supplementary Table~S9). On the medium-scale transfer study, \sgm{} also improves completed tasks by $7.3 \pm 3.3\%$, indicating that the controller remains beneficial outside the primary small-map benchmark (Supplementary Table~S10).

\subsection{Endpoint steering drives throughput; route guidance improves execution}

The component ablations identify the roles of the two SGM interfaces. Endpoint-only retains memory-guided endpoint scoring but uses unit routing costs. It matches full \sgm{} in final completed tasks across the small-map ablation benchmark: averaged over the 15 map--fleet conditions, endpoint-only completes 6395.2 tasks and full \sgm{} completes 6394.3. This is a direct mechanistic result, not a weakness of the method: it establishes endpoint instantiation as the primary throughput-control interface.

Route guidance improves the quality and cost of execution while preserving that throughput. Relative to endpoint-only, full \sgm{} lowers total planner time in every ablation condition, reduces waiting in 13 of 15 conditions, and reduces blocked moves and blocked-motion replans in 12 conditions. Thus, endpoint steering determines which work enters the planner, while route guidance reduces the execution burden of serving that work.

The remaining ablations reinforce this decomposition. Routing-only performs substantially worse than controllers with endpoint memory, showing that routing cannot recover from poor endpoint instantiation after goals are fixed. Removing queue preservation also reduces completed tasks and increases congestion, because repeated reconstruction discards useful allocation continuity. Together, these results show that SGM's throughput gain is produced by memory-guided endpoint steering, while its route guidance and queue-preserving execution policy make that throughput more efficient to plan and execute.

\asgm{} serves a different operating point. It reduces allocation overhead relative to the full controller, but sacrifices absolute completed-task throughput. We therefore treat \asgm{} as an adaptive compute--throughput trade-off, whereas \sgm{} is the primary controller for the main performance claims.

\begin{table}[h]
\centering
\caption{Full \sgm{} versus endpoint-only, averaged over the 15 small-map map--fleet conditions. Lower is better for planner time, waiting, blocked moves, and blocked-motion replans.}
\label{tab:routing-value}
\scriptsize
\begin{tabular}{lrrr}
\toprule
Metric & Endpoint-only & \sgm{} & Change \\
\midrule
Final completed tasks & 6395.2 & 6394.3 & $-0.9$ \\
Planner time (s) & 295.6 & 268.8 & $-9.1\%$ \\
Waiting events / step & 27.24 & 27.03 & $-0.8\%$ \\
Blocked moves & 19.1 & 16.9 & $-11.5\%$ \\
Blocked-motion replans & 8.84 & 7.90 & $-10.6\%$ \\
\bottomrule
\end{tabular}
\end{table}

\section{Discussion}
\label{sec:discussion}

\paragraph{Endpoint instantiation is the primary throughput-control interface.}
The central result is that many-to-many MAPD congestion can be shaped before a path is planned. \sgm{} uses recent execution history to avoid repeatedly committing new tasks to source--destination combinations associated with waiting, blocking, and concentrated traffic. The endpoint-only ablation identifies this as the dominant throughput mechanism: it retains 99.3--100.5\% of full \sgm{} throughput across the evaluated small-map conditions. Thus, most of the performance gain comes from reading graph memory when feasible endpoints are instantiated, rather than only after goals have been fixed.

\paragraph{Why the complete controller retains route guidance.}
Endpoint-only is deliberately a throughput-focused ablation, not the recommended deployment controller. Although it matches full \sgm{} in final completed tasks, route guidance improves the operational cost of achieving that throughput. Across the 15 small-map map--fleet conditions, full \sgm{} reduces planner time by 9.1\% in every condition, reduces waiting events in 13 conditions, and reduces blocked moves and blocked-motion replans in 12 conditions. These effects are consistent with the intended role of route guidance: after endpoint steering selects where work should be served, directed-edge memory helps the planner serve that work with less contention. Endpoint steering controls which goals enter the planner; route guidance reduces the execution and planning burden of serving them.

\paragraph{Controller choice and computational cost.}
\sgm{} is the primary controller when absolute completed work is the objective: it has the highest completed-task count at every fleet level. \asgm{} provides a distinct adaptive operating point. It limits allocation computation under high load and achieves the strongest compute-normalized throughput at the high fleet level, while retaining a substantial improvement over both M2M baselines. Thus, \sgm{} is the appropriate choice when throughput is paramount, whereas \asgm{} is useful when allocation computation is a tighter operational constraint.

\begin{table}[t]
\centering
\caption{Throughput--compute trade-off over five layouts and 25 seeds per condition. \emph{/M cand.} is aggregate completed tasks divided by aggregate candidate evaluations in millions; \emph{/comp. s} is aggregate completed tasks divided by aggregate allocator wall-clock seconds.}
\label{tab:efficiency}
\scriptsize
\begin{tabular*}{\columnwidth}{@{\extracolsep{\fill}}llrrr@{}}
\toprule
Fleet & Method & Completed & /M cand. & /comp. s \\
\midrule
Low & M2M & 3893.9 & 1.96 & 3.07 \\
& M2M-wSKU & 4168.5 & 2.96 & 4.01 \\
& \asgm{} & 4207.7 & 2.51 & 5.13 \\
& \sgm{} & \textbf{5539.7} & \textbf{4.11} & \textbf{6.00} \\
\midrule
Medium & M2M & 4614.2 & 1.54 & 2.28 \\
& M2M-wSKU & 4950.4 & 2.42 & 3.19 \\
& \asgm{} & 5039.4 & 2.09 & 4.31 \\
& \sgm{} & \textbf{6615.1} & \textbf{3.22} & \textbf{4.76} \\
\midrule
High & M2M & 5157.9 & 1.03 & 1.46 \\
& M2M-wSKU & 5591.7 & 1.62 & 2.10 \\
& \asgm{} & 6113.0 & \textbf{3.79} & \textbf{5.85} \\
& \sgm{} & \textbf{7027.9} & 2.08 & 3.14 \\
\bottomrule
\end{tabular*}
\end{table}

\paragraph{Robustness, controls, and transfer.}
The control studies distinguish SGM from two common routing-only alternatives. Adding recent routing costs or a static highway after M2M-wSKU has fixed endpoints leaves throughput near the M2M-wSKU baseline, whereas \sgm{} gains substantially on the same cells. Matched-candidate-cap studies show that the result is not explained by evaluating more endpoint pairs, and one-factor sweeps show that the improvement persists across a broad range of memory settings. The medium open and open-top transfer study further shows that \sgm{} remains beneficial beyond the primary small-map benchmark.

\paragraph{Scope of the claim.}
The evidence establishes SGM as an effective congestion-control layer for many-to-many warehouse MAPD under shared rolling-horizon RHCR/PBS planning. These results support endpoint instantiation as a robust control interface for shaping congestion before planner goals are fixed. Future work includes nonstationary demand, richer kinematics, and physical robot deployments.

\section{Conclusion}
\label{sec:conclusion}

We introduced \sgm{}, a bounded graph-memory layer for many-to-many MAPD that uses recent execution to guide endpoint instantiation and route preferences. Across paired benchmarks, \sgm{} improves completed-task throughput in every evaluated map--fleet condition relative to both reconstructed many-to-many baselines. The endpoint-only ablation shows that memory-guided endpoint instantiation is the primary source of this gain.

The complete \sgm{} controller retains this throughput while reducing waiting, blocked motion, blocked-motion replanning, and planner runtime. These results show that many-to-many MAPD offers a high-leverage congestion-control interface before path planning begins: the controller can shape which goals enter the planner, not only how agents travel to goals already selected. Future work includes extending SGM to nonstationary warehouse environments and physical robot deployments.

\bibliography{aaai2027}


\let\bibliography\arxivbibliography
\clearpage
\bibliography{aaai2027}

\clearpage
\renewcommand{\bibliography}[1]{}
\let\arxivappendix\appendix
\renewcommand{\appendix}{%
  \arxivappendix
  \setcounter{section}{0}%
  \setcounter{figure}{0}%
  \setcounter{table}{0}%
  \setcounter{equation}{0}%
  \renewcommand{\thesection}{S\arabic{section}}%
  \renewcommand{\thefigure}{S\arabic{figure}}%
  \renewcommand{\thetable}{S\arabic{table}}%
  \renewcommand{\theequation}{S\arabic{equation}}%
}
\def\maketitle{}

\maketitle

\noindent\textbf{Purpose.} This technical supplement provides the complete
experimental record supporting the paper: controller definitions, memory
configuration, feasibility preservation, full benchmark outcomes, ablations,
robustness controls, transfer evidence, and spatial diagnostics.

\appendix

\section{Reproducibility Protocol and Statistical Analysis}
\label{app:protocol}

The primary benchmark comprises $5\times3\times25\times4=1500$
method--layout--fleet--seed runs. The five layouts are restricted, open-top,
open, maze~A, and maze~B. Low, medium, and high fleet levels use 28, 56, and
84 agents, respectively; these are 30.1\%, 60.2\%, and 90.3\% of the 93
designated parking cells. Every run lasts $H=3600$ timesteps. Within each
layout--fleet--seed cell, every method receives the same map, initial state, and
replayed request stream.

The primary outcome is the cumulative number of completed tasks, $C(H)$; final
throughput is $C(H)/H$. We also record time-resolved completions, final-window
activity, blocked moves, waiting events, planner time, allocation time, service
time, and candidate evaluations. All main comparisons use the same RHCR/PBS
backend with a planning window of 512. The allocator sequence depth is three;
the 30-second allocator and 10-second planner limits are safety budgets.

\paragraph{Computing environment.} Experiments ran on a cluster node that has two Intel Xeon Gold 6248 processors (48 physical
cores and 96 hardware threads total).
The experiment campaign used 88 CPU threads and a 1~TB memory allocation; the
simulation and planner are CPU-bound, so no GPU was used. Recorded runs used
Python~3.10.12 on Linux~5.15.0-179-generic (glibc~2.35), with GCC~11.4.0 for
the native planner extension. The Python environment pins PyYAML~6.0.3,
pybind11~3.0.4, NumPy~2.4.6, SciPy~1.17.1, pandas~3.0.3, Matplotlib~3.11.0,
and statsmodels~0.14.6.

For each layout--fleet condition, we compare seed-matched final completed-task
counts using two-sided Wilcoxon signed-rank tests. Holm--Bonferroni correction
controls the family-wise error rate across the 30 planned comparisons of \sgm{}
with M2M and M2M-wSKU. Table~\ref{tab:wilcoxon} gives the complete test record.
All adjusted $p$-values are below $6.14\times10^{-5}$.

\paragraph{Paired inference.} Pairing makes the seed-level throughput
difference, rather than an unpaired aggregate, the unit of inference. Each
comparison therefore holds the layout, fleet size, initial configuration, and
entire request sequence fixed while varying only the controller. The reported
paired percentage gains summarize effect magnitude, and the signed-rank tests
establish that the observed advantages are consistently positive across the
replayed streams.

\begin{anchoredblock}
\captionof{table}{Two-sided Wilcoxon signed-rank tests for final completed
tasks ($n=25$ paired seeds per cell). Each comparison is \sgm{} versus the
named baseline; $p_{\mathrm{Holm}}$ is adjusted across all 30 planned tests.}
\label{tab:wilcoxon}
\small
\setlength{\tabcolsep}{8pt}
\resizebox{\columnwidth}{!}{%
\begin{tabular}{llrrrr}
\toprule
Layout & Fleet & $W_{\mathrm{M2M}}$ & $p_{\mathrm{Holm}}$ &
  $W_{\mathrm{wSKU}}$ & $p_{\mathrm{Holm}}$ \\
\midrule
Restricted & Low    & 0  & $1.79\times10^{-6}$ & 0 & $1.79\times10^{-6}$ \\
Restricted & Medium & 0  & $1.79\times10^{-6}$ & 0 & $1.79\times10^{-6}$ \\
Restricted & High   & 1  & $1.79\times10^{-6}$ & 0 & $1.79\times10^{-6}$ \\
Open-top   & Low    & 0  & $1.79\times10^{-6}$ & 0 & $1.79\times10^{-6}$ \\
Open-top   & Medium & 15 & $4.90\times10^{-5}$ & 0 & $1.79\times10^{-6}$ \\
Open-top   & High   & 17 & $6.14\times10^{-5}$ & 0 & $1.79\times10^{-6}$ \\
Open       & Low    & 0  & $1.79\times10^{-6}$ & 0 & $1.79\times10^{-6}$ \\
Open       & Medium & 1  & $1.79\times10^{-6}$ & 0 & $6.14\times10^{-5}$ \\
Open       & High   & 0  & $1.79\times10^{-6}$ & 2 & $6.14\times10^{-5}$ \\
Maze A     & Low    & 0  & $6.14\times10^{-5}$ & 0 & $6.14\times10^{-5}$ \\
Maze A     & Medium & 0  & $1.79\times10^{-6}$ & 0 & $1.79\times10^{-6}$ \\
Maze A     & High   & 0  & $1.79\times10^{-6}$ & 0 & $1.79\times10^{-6}$ \\
Maze B     & Low    & 0  & $1.79\times10^{-6}$ & 0 & $1.79\times10^{-6}$ \\
Maze B     & Medium & 0  & $1.79\times10^{-6}$ & 0 & $1.79\times10^{-6}$ \\
Maze B     & High   & 0  & $1.79\times10^{-6}$ & 3 & $2.09\times10^{-6}$ \\
\bottomrule
\end{tabular}
}
\end{anchoredblock}

\section{Controllers and Common Evaluation Harness}
\label{app:controllers}

M2M and M2M-wSKU are released-code-faithful reconstructions of the allocation
objectives in \citet{schneider2026manytomany}, integrated into the common
simulator, request-replay protocol, RHCR/PBS backend, and validation layer.
This controlled integration isolates the allocation decision under one planner
boundary. M2M scores candidates from travel-distance estimates. M2M-wSKU adds the released
inventory-distribution term, with base travel weight $w_b=1.0$ and SKU weight
$w_s=0.25$. Both reconstructions rebuild queued tasks at each allocation cycle.

\sgm{} is the primary controller. It continuously applies path-memory endpoint
scoring, supplies bounded memory-derived route costs, and preserves queued work
across event-driven repairs. \asgm{} uses the same memory representation but
activates memory-guided scoring and routing only under sustained congestion
evidence; otherwise it uses the M2M-wSKU-style objective and unit route costs.
Table~\ref{tab:policies} summarizes the four main methods.

\begin{anchoredblock}
\captionof{table}{Controller implementation details under the common evaluation
harness. All methods use the same simulator, replayed requests, RHCR/PBS backend,
and feasibility validation.}
\label{tab:policies}
\scriptsize
\setlength{\tabcolsep}{5pt}
\resizebox{\columnwidth}{!}{%
\begin{tabular}{lp{1.45in}lll}
\toprule
Method & Endpoint objective & Route costs & Memory activation & Queue update \\
\midrule
M2M & Travel distance & Unit & None & Rebuild \\
M2M-wSKU & $w_b d + w_s I$; $w_b=1.0$, $w_s=0.25$ & Unit & None & Rebuild \\
\asgm{} & M2M-wSKU or memory-scored & Unit or bounded memory-derived & Sustained congestion & Preserve \\
\sgm{} & Path-memory scored & Bounded memory-derived & Continuous & Preserve \\
\bottomrule
\end{tabular}
}
\end{anchoredblock}

\section{Memory Configuration and Validity Boundary}
\label{app:memory}

\subsection{Execution traces and deposits}

\sgm{} maintains typed, decaying signals rather than a single congestion value.
Node channels record waiting, endpoint pressure, and completion. Directed-edge
channels record traversal, delay, blocking, and directional flow. When a request
is released, one unit of source or destination pressure is distributed uniformly
over its feasible endpoints. Unplanned waiting deposits waiting and delay;
planned waiting on a valid route does not. Before execution, proposed moves are
checked for vertex conflicts and edge swaps. A move that would violate either
constraint is held and deposits a \emph{pre-execution blocking} signal; it is not
an executed collision.

For channel $c$ and graph element $x$, memory evolves as
\begin{equation}
M_c^{t+1}(x)=\rho_c M_c^t(x)+\Delta_c^t(x).
\label{eq:decay}
\end{equation}
Waiting and delay use $\rho_c=0.85$; blocking and endpoint pressure use 0.90;
congestion, traversal, and flow use 0.92; and completion uses 0.95. A corridor
modifier can increase retention by 0.05 on degree-at-most-two corridors, capped
at 0.99.

\begin{anchoredblock}
\captionof{table}{Active memory channels in the reported configuration.}
\label{tab:channels}
\small
\setlength{\tabcolsep}{9pt}
\resizebox{\columnwidth}{!}{%
\begin{tabular}{llll}
\toprule
Channel & Element & Deposit event & Retention \\
\midrule
Traversal / successful flow & Directed edge & Executed traversal & .92 \\
Waiting & Node & Unplanned wait & .85 \\
Delay & Directed edge & Unplanned wait & .85 \\
Blocking & Node / directed edge & Blocked proposed move & .90 \\
Endpoint pressure & Node & Request release & .90 \\
Completion & Endpoint node & Completed delivery & .95 \\
\bottomrule
\end{tabular}
}
\end{anchoredblock}

\subsection{Endpoint and routing interfaces}

For a feasible candidate $(a_i,\tau,s,d)$, \sgm{} augments the M2M-wSKU-style
baseline score with sampled path memory:
\begin{equation}
C(a_i,\tau,s,d)=C_0(a_i,\tau,s,d)+1.75P_{\mathrm{path}}(a_i,s,d).
\label{eq:endpoint-score-app}
\end{equation}
The path term sums current directed penalties along sampled unit-cost shortest
paths from $a_i$ to $s$ and from $s$ to $d$. Source and destination shortlists
are capped at 64 locations each, with at most 32 endpoint pairs per request.
These sampled paths are scoring proxies: RHCR/PBS computes the final joint plan.

For a legal directed move $(u,v)$, routing uses
\begin{equation}
\widetilde c_t(u,v)=1+\min\{0.1,\;0.2R_t(u,v)\},
\label{eq:routing-cost-app}
\end{equation}
where $R_t$ aggregates normalized congestion, blocking, delay, contraflow,
successful forward flow, and discounted planned occupancy over 32 timesteps.
Thus every legal move has positive, bounded cost:
$1\leq\widetilde c_t(u,v)\leq1.1$.

\begin{anchoredblock}
\captionof{table}{Remaining reported controller settings.}
\label{tab:settings}
\small
\setlength{\tabcolsep}{18pt}
\resizebox{\columnwidth}{!}{%
\begin{tabular}{lr}
\toprule
Setting & Value \\
\midrule
Planning horizon & 3600 timesteps \\
Paired seeds & 25 (10--34) \\
RHCR/PBS window & 512 \\
Allocation sequence depth & 3 \\
Path-memory weight & 1.75 \\
Routing-memory weight / cap & 0.2 / 0.1 \\
Guidance horizon & 32 \\
\sgm{} source / destination / pair caps & 64 / 64 / 32 \\
M2M source / destination / pair caps & 100 / 100 / 128 \\
\asgm{} fallback pair cap & 160 \\
\bottomrule
\end{tabular}
}
\end{anchoredblock}

\subsection{Feasibility preservation}

\paragraph{Proposition.} For every endpoint assignment selected by \sgm{}, the
RHCR/PBS backend receives the original graph, agent states, task-feasibility
constraints, and vertex- and edge-conflict predicates. Every accepted and
executed plan is therefore collision-free and task-feasible under the original
MAPD constraints.

\paragraph{Justification.} \sgm{} changes only scalar scores used to rank a
finite set of feasible candidates and scalar costs assigned to legal moves. It
does not add or remove graph edges, alter agent states, relax endpoint
feasibility, or modify conflict predicates. RHCR/PBS alone accepts the joint
plan. Positive bounded route costs and finite candidate ranking cannot create an
infeasible assignment, a negative edge cost, a zero-cost cycle, or an infinite
cost for a finite legal path.

\clearpage
\onecolumn
\section{Layouts and Complete Benchmark Outcomes}
\label{app:benchmark}

Each layout is a $50\times27$ grid with 1350 cells and 93 designated
storage/parking cells. Restricted, open-top, and open are the warehouse layouts
used for the primary benchmark; maze~A and maze~B preserve the scale while
changing corridor structure.

\begin{anchoredblock}
\includegraphics[width=1\textwidth]{figures/map_layouts_compact.pdf}
\captionof{figure}{The five $50\times27$ primary-benchmark layouts at full scale. Dark
cells are obstacles; light cells are traversable aisles; colored cells denote
inbound stations, outbound stations, and storage/parking locations. Restricted,
open-top, and open are the benchmark warehouse families; maze~A and maze~B
hold scale fixed while changing corridor topology.}
\label{fig:layouts}
\end{anchoredblock}

\begin{anchoredblock}
\captionof{table}{Cell composition of the five layouts (counts out of 1350).}
\label{tab:cells}
\small
\setlength{\tabcolsep}{12pt}
\resizebox{0.70\columnwidth}{!}{%
\begin{tabular}{lrrrrr}
\toprule
Layout & Obstacle & Aisle & Inbound & Outbound & Storage \\
\midrule
Restricted & 550 & 207 & 350 & 150 & 93 \\
Open-top   & 500 & 307 & 300 & 150 & 93 \\
Open       & 375 & 382 & 350 & 150 & 93 \\
Maze A     & 408 & 349 & 350 & 150 & 93 \\
Maze B     & 390 & 367 & 350 & 150 & 93 \\
\bottomrule
\end{tabular}
}
\end{anchoredblock}

Table~\ref{tab:full-results} gives all primary-benchmark terminal outcomes.
The final column is the paired \sgm{} percentage improvement over the stronger
M2M or M2M-wSKU baseline in the corresponding cell.

\begin{anchoredblock}
\captionof{table}{Complete primary benchmark: final completed tasks (mean over
25 paired seeds). Gain is the paired \sgm{} improvement over the stronger M2M
baseline, with a 95\% confidence interval.}
\label{tab:full-results}
\small
\setlength{\tabcolsep}{7pt}
\resizebox{0.70\columnwidth}{!}{%
\begin{tabular}{llrrrrr}
\toprule
Layout & Fleet & M2M & M2M-wSKU & \asgm{} & \sgm{} & Gain (\%) \\
\midrule
Restricted & Low    & 4275 & 4610 & 4648 & \textbf{6123} & $33.2\pm5.5$ \\
Restricted & Medium & 5063 & 5542 & 5628 & \textbf{7551} & $36.7\pm6.9$ \\
Restricted & High   & 5746 & 6280 & 6955 & \textbf{7821} & $24.3\pm4.8$ \\
Open-top   & Low    & 4221 & 4313 & 4346 & \textbf{5770} & $34.6\pm6.7$ \\
Open-top   & Medium & 5062 & 5016 & 5255 & \textbf{6878} & $34.9\pm11.0$ \\
Open-top   & High   & 5573 & 5420 & 6156 & \textbf{7243} & $28.8\pm9.3$ \\
Open       & Low    & 4126 & 4386 & 4423 & \textbf{5820} & $33.2\pm4.6$ \\
Open       & Medium & 4920 & 5269 & 5347 & \textbf{6949} & $32.5\pm5.4$ \\
Open       & High   & 5595 & 6287 & 6512 & \textbf{7517} & $20.5\pm5.8$ \\
Maze A     & Low    & 2899 & 3258 & 3323 & \textbf{4425} & $36.1\pm2.9$ \\
Maze A     & Medium & 3346 & 3866 & 3898 & \textbf{5149} & $33.4\pm3.3$ \\
Maze A     & High   & 3843 & 4367 & 4862 & \textbf{5791} & $32.7\pm4.2$ \\
Maze B     & Low    & 3949 & 4275 & 4299 & \textbf{5560} & $30.4\pm4.4$ \\
Maze B     & Medium & 4681 & 5059 & 5069 & \textbf{6548} & $29.6\pm5.3$ \\
Maze B     & High   & 5033 & 5604 & 6080 & \textbf{6768} & $21.1\pm4.8$ \\
\bottomrule
\end{tabular}
}
\end{anchoredblock}

\twocolumn
\begin{figure*}[!t]
\centering
\includegraphics[width=0.99\textwidth]{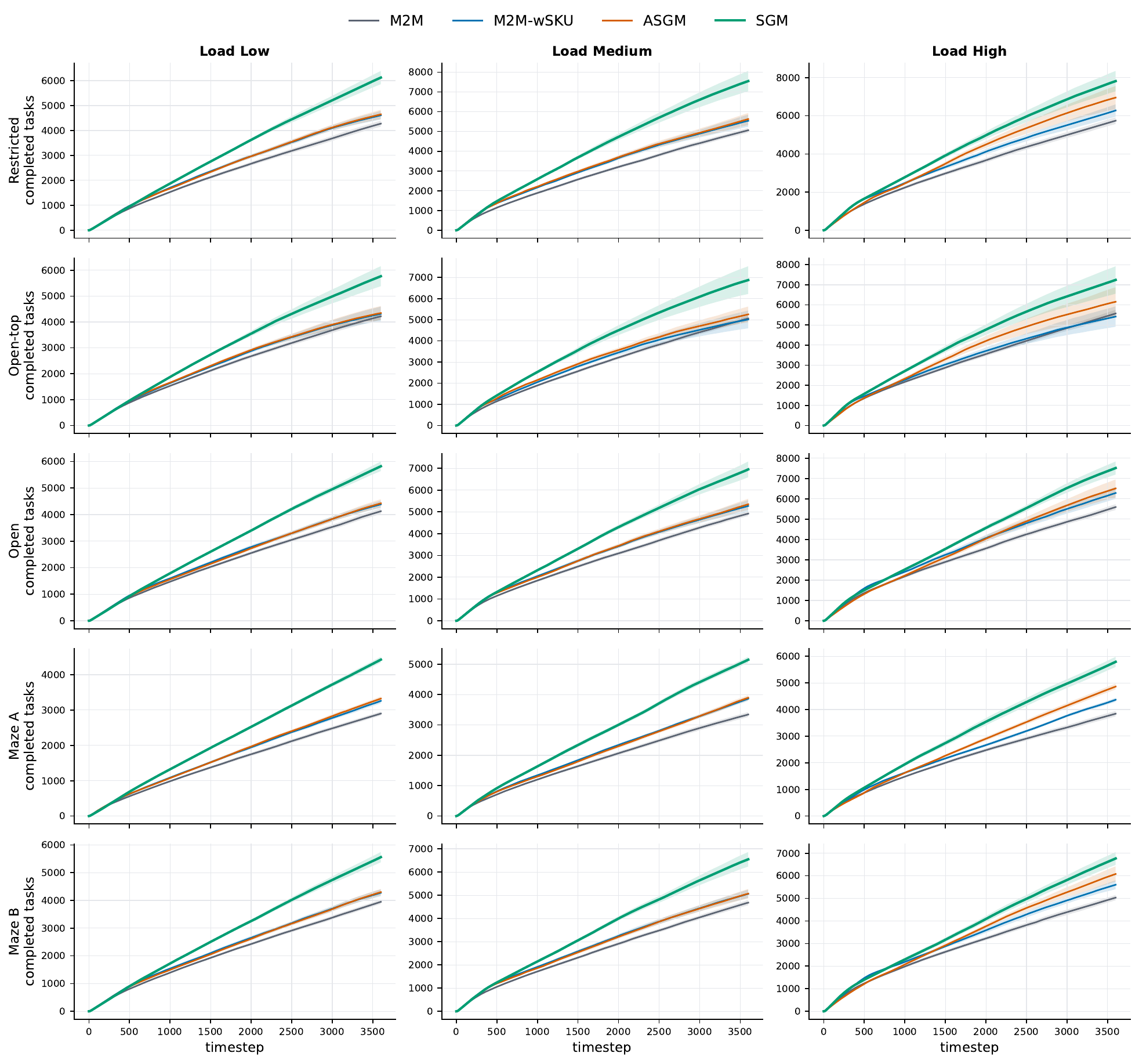}
\caption{Cumulative completed tasks throughout the 3600-timestep horizon for
every primary-benchmark layout and fleet level. Curves are means over 25 paired
seeds; shaded bands show 95\% confidence intervals. SGM separates from both
M2M baselines early and maintains the advantage through the terminal horizon in
all 15 conditions.}
\label{fig:all-throughput}
\end{figure*}

\FloatBarrier
\paragraph{Temporal separation.} The trajectories show that SGM's advantage
is not a terminal-window artifact. The methods begin near parity while agents
first receive and instantiate requests, then the SGM curves separate as the
shared system accumulates traffic. That separation is sustained through the
3600-timestep horizon in every layout--fleet condition, including the two maze
topologies. The result is consistent with the proposed mechanism: endpoint
selection changes the congestion pattern before repeated assignments commit
traffic to the same constrained regions.

\section{Ablations and Robustness Controls}
\label{app:controls}

Endpoint-only retains path-memory endpoint steering with unit routing costs.
Routing-only retains memory-weighted routing without endpoint-memory scoring.
No-queue-preserve keeps both memory interfaces but rebuilds queued work at every
allocation cycle. The ablation cleanly identifies endpoint steering as the
dominant throughput interface: endpoint-only retains 99.3--100.5\% of full
\sgm{} throughput across fleet levels. Table~\ref{tab:ablation} and
Figure~\ref{fig:ablation-summary} report the same conclusion in tabular and
visual form.

\begin{anchoredblock}
\captionof{table}{Component ablations aggregated over five layouts and 25 seeds
per fleet level. ``\% SGM'' is relative to full \sgm{} throughput; $\Delta$
wSKU is relative to M2M-wSKU.}
\label{tab:ablation}
\small
\setlength{\tabcolsep}{2pt}
\renewcommand{\arraystretch}{1.08}
\begin{tabular}{@{}p{0.45in}p{0.93in}rrr@{}}
\toprule
Fleet & Method & Completed & \% \sgm{} & $\Delta$ wSKU \\
\midrule
Low & \sgm{} & 5540 & 100.0 & $+32.9$ \\
    & Endpoint-only & 5556 & 100.3 & $+33.3$ \\
    & Routing-only & 4347 & 78.5 & $+4.3$ \\
    & No queue preserve & 5326 & 96.1 & $+27.8$ \\
Medium & \sgm{} & 6615 & 100.0 & $+33.6$ \\
    & Endpoint-only & 6570 & 99.3 & $+32.7$ \\
    & Routing-only & 4968 & 75.1 & $+0.4$ \\
    & No queue preserve & 6460 & 97.6 & $+30.5$ \\
High & \sgm{} & 7028 & 100.0 & $+25.7$ \\
    & Endpoint-only & 7060 & 100.5 & $+26.3$ \\
    & Routing-only & 5537 & 78.8 & $-1.0$ \\
    & No queue preserve & 6913 & 98.4 & $+23.6$ \\
\bottomrule
\end{tabular}
\renewcommand{\arraystretch}{1}
\end{anchoredblock}

\begin{anchoredblock}
\includegraphics[width=0.85\columnwidth]{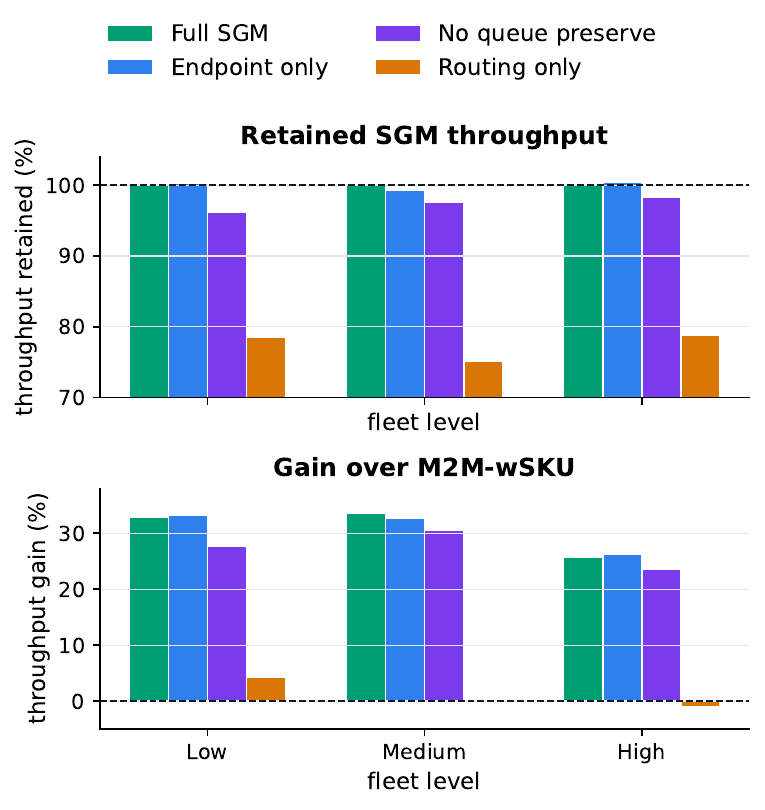}
\captionof{figure}{Component attribution aggregated over the five primary layouts and
25 paired seeds per fleet level. Endpoint-only preserves full SGM throughput,
whereas routing-only does not; routing and queue preservation contribute the
operational efficiency retained by the full controller.}
\label{fig:ablation-summary}
\end{anchoredblock}

\paragraph{Operational value of route guidance.} Endpoint-only isolates the
throughput effect of endpoint steering. Full \sgm{} retains that throughput while
making its execution less costly: it lowers planner time in every small-map
condition and reduces waiting, blocked moves, and blocked-motion replans in most
conditions (Table~\ref{tab:routing-value-app}).

\begin{anchoredblock}
\captionof{table}{Full \sgm{} versus endpoint-only, averaged over the 15
small-map layout--fleet conditions. Lower is better except for completed tasks.}
\label{tab:routing-value-app}
\small
\setlength{\tabcolsep}{9pt}
\resizebox{\columnwidth}{!}{%
\begin{tabular}{lrrrr}
\toprule
Metric & Endpoint-only & \sgm{} & Change & \sgm{} favored \\
\midrule
Final completed tasks & 6395.2 & 6394.3 & $-0.9$ & -- \\
Planner time (s) & 295.6 & 268.8 & $-9.1\%$ & 15 / 15 \\
Waiting events / step & 27.24 & 27.03 & $-0.8\%$ & 13 / 15 \\
Blocked moves & 19.1 & 16.9 & $-11.5\%$ & 12 / 15 \\
Blocked-motion replans & 8.84 & 7.90 & $-10.6\%$ & 12 / 15 \\
\bottomrule
\end{tabular}
}
\end{anchoredblock}

Routing controls hold M2M-wSKU endpoint allocation fixed and vary only the
routing costs. They show that post-commitment route guidance alone does not
produce the main effect. Sensitivity controls vary one setting at a time;
matched-cap controls reverse the default candidate caps. Table~\ref{tab:controls}
gives the complete control record on restricted, open-top, and open layouts at
medium and high fleet levels (five paired seeds per cell).

\clearpage
\onecolumn
\begin{anchoredblock}
\captionof{table}{Complete routing, sensitivity, and candidate-cap controls.
Each row pools 30 matched layout--fleet--seed runs. $\Delta$ is the paired
percentage change relative to M2M-wSKU.}
\label{tab:controls}
\small
\setlength{\tabcolsep}{6pt}
\renewcommand{\arraystretch}{1.08}
\begin{tabular*}{\columnwidth}{@{\extracolsep{\fill}}llrr@{}}
\toprule
Control & Setting & Completed tasks & $\Delta$ (\%) \\
\midrule
\multicolumn{4}{@{}l}{\textit{Routing controls: M2M-wSKU endpoint allocation is held fixed.}} \\
M2M-wSKU & Unit routing costs & $5749.6\pm326.3$ & $0.0\pm0.0$ \\
M2M-wSKU & Recent-traffic routing & $5737.3\pm328.6$ & $-0.1\pm2.1$ \\
M2M-wSKU & Static highway routing & $5661.5\pm338.3$ & $-1.4\pm2.5$ \\
\sgm{} & Endpoint and route memory & \textbf{$7391.1\pm484.3$} & \textbf{$+29.2\pm5.7$} \\
\addlinespace[1pt]
\multicolumn{4}{@{}l}{\textit{One-factor sensitivity: full \sgm{} with one setting varied.}} \\
Path-memory weight & $.875$ & $7427.7\pm496.9$ & $+29.7\pm5.6$ \\
                   & $1.75$ (reported) & $7391.1\pm484.3$ & $+29.2\pm5.7$ \\
                   & $3.5$ & $7384.5\pm499.6$ & $+29.1\pm6.1$ \\
Memory retention & $.90$ & $7400.4\pm505.4$ & $+29.2\pm5.9$ \\
                 & $.95$ & $7381.2\pm496.0$ & $+28.9\pm5.6$ \\
                 & $.98$ & $7351.0\pm492.0$ & $+28.5\pm5.8$ \\
Endpoint-pair cap & $16$ & $6603.4\pm360.7$ & $+16.0\pm5.2$ \\
                  & $32$ (reported) & $7391.1\pm484.3$ & $+29.2\pm5.7$ \\
                  & $64$ & $8084.8\pm671.7$ & $+40.5\pm7.5$ \\
Route-memory weight & $0.0$ & $7582.8\pm507.9$ & $+32.5\pm5.9$ \\
                    & $0.2$ (reported) & $7391.1\pm484.3$ & $+29.2\pm5.7$ \\
                    & $0.4$ & $7411.4\pm501.0$ & $+29.4\pm5.7$ \\
\addlinespace[1pt]
\multicolumn{4}{@{}l}{\textit{Matched candidate caps: source / destination / pair caps are exchanged.}} \\
M2M-wSKU & $64/64/32$ & $5687.7\pm321.7$ & $-0.9\pm1.9$ \\
\sgm{} & $100/100/128$ & $8351.8\pm754.6$ & $+44.8\pm8.5$ \\
\bottomrule
\end{tabular*}
\renewcommand{\arraystretch}{1}
\end{anchoredblock}

\section{Transfer and Spatial Evidence}
\label{app:transfer}

The medium-scale transfer study uses open and open-top layouts at the high fleet
level, five paired seeds per layout, and the same 3600-timestep horizon. \sgm{}
improves final completed tasks over M2M-wSKU by $7.3\pm3.3\%$. The transfer
result confirms that the endpoint-memory interface remains effective beyond the
primary small layouts.

\begin{anchoredblock}
\captionof{table}{Medium-scale high-fleet transfer results on open and open-top
layouts (10 paired runs: two layouts by five seeds). $\Delta$ is relative to
M2M-wSKU.}
\label{tab:transfer-results}
\small
\setlength{\tabcolsep}{12pt}
\resizebox{0.73\columnwidth}{!}{%
\begin{tabular}{lrrrr}
\toprule
Method & $n$ & Completed tasks & $\Delta$ (\%) & Range (\%) \\
\midrule
M2M-wSKU & 10 & $25764.6\pm1204.2$ & $0.0\pm0.0$ & $0.0$ to $0.0$ \\
Recent routing & 10 & $25490.4\pm1310.0$ & $-1.1\pm0.7$ & $-3.0$ to $1.4$ \\
Static highway & 10 & $24828.6\pm1233.8$ & $-3.6\pm1.5$ & $-8.6$ to $-0.6$ \\
Endpoint-only & 10 & $23150.9\pm5040.5$ & $-8.6\pm21.0$ & $-77.4$ to $17.9$ \\
\sgm{} & 10 & $27576.0\pm795.9$ & $+7.3\pm3.3$ & $+1.4$ to $14.8$ \\
\bottomrule
\end{tabular}
}
\end{anchoredblock}

The transfer comparison makes the role of the full controller explicit.
Endpoint-only is competitive on the primary small maps, but it does not retain
the full controller's performance on these larger layouts. The routing-only
controls remain near or below M2M-wSKU in both benchmark regimes, whereas full
\sgm{} remains positive under the changed scale and geometry.

\clearpage
\twocolumn[
\begin{center}
\resizebox{0.98\textwidth}{!}{\includegraphics{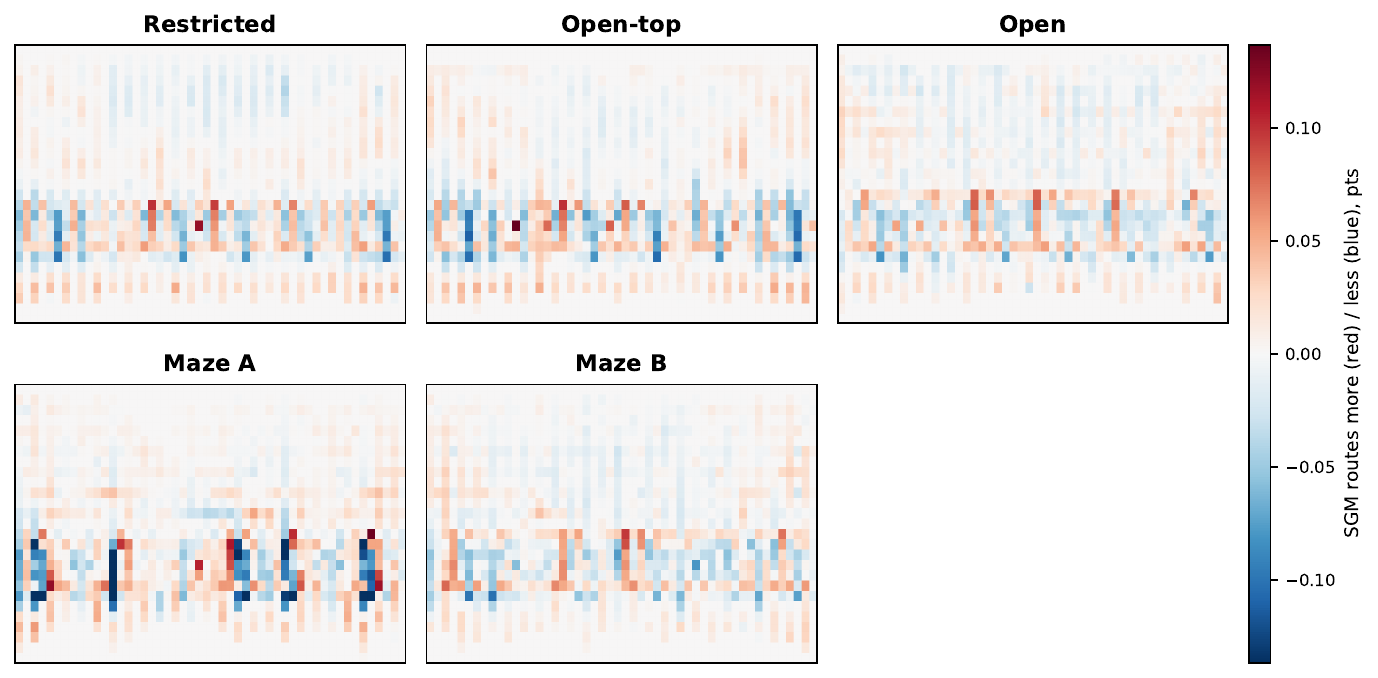}}
\captionof{figure}{Change in final-window traversal share at high fleet level:
SGM minus M2M-wSKU. Red cells receive a larger share of traversals under SGM;
blue cells receive a smaller share. Each panel is normalized by its method's
total traversals, so the diagnostic measures redistribution rather than the
higher absolute task volume produced by SGM.}
\label{fig:spatial-traffic}
\end{center}
]

\subsection{Spatial traffic redistribution}

Figure~\ref{fig:spatial-traffic} complements terminal counts with a
traffic-distribution diagnostic at high fleet level. It visualizes the change
in each cell's share of traversals relative to M2M-wSKU during the final
measurement window. The redistribution is topology-specific: SGM redirects
traffic away from the locally overloaded regions induced by each layout.

The diagnostic does not require any particular corridor to receive more or less
traffic. Instead, it shows that the redistribution follows each layout's local
geometry: under \sgm{}, relative traversal decreases in some congested service
regions and increases on alternative accessible routes. Together with the
transfer table, this evidence supports a controller that responds to the
available warehouse structure rather than a fixed global traffic pattern.

\bibliography{aaai2027}

\end{document}